# Decoding the mechanisms underlying cell-fate decision-making during stem cell differentiation by Random Circuit Perturbation

Bin Huang[1,*], Mingyang Lu[2,*], Madeline Galbraith[1,3], Herbert Levine[1,4,5,#], Jose N. Onuchic[1,3,6,7,#], Dongya Jia[1,#]

[1]Center for Theoretical Biological Physics, Rice University, Houston, TX 77005, USA
[2]The Jackson Laboratory, 600 Main St, Bar Harbor, ME 04609, USA
[3]Department of Physics and Astronomy, Rice University, Houston, TX 77005, USA
[4]Department of Bioengineering, Northeastern University, Boston, MA 02115, USA
[5]Department of Physics, Northeastern University, Boston, MA 02115, USA
[6]Department of Chemistry, Rice University, Houston, TX 77005, USA
[7]Department of Bioengineering, Rice University, Houston, TX 77005, USA
*These authors contributed equally to this work.
#To whom correspondence can be addressed:
D.J. (dj9@rice.edu), H.L. (h.levine@northeastern.edu) and J.N.O. (jonuchic@rice.edu).

**Abstract**

Stem cells can precisely and robustly undergo cellular differentiation and lineage commitment, referred to as stemness. However, how the gene network underlying stemness regulation reliably specifies cell fates is not well understood. To address this question, we applied a recently developed computational method, *Ra*ndom *Ci*rcuit *Pe*rturbation (RACIPE), to a nine-component gene regulatory network (GRN) governing stemness, from which we identified fifteen robust gene states. Among them, four out of the five most probable gene states exhibit gene expression patterns observed in single mouse embryonic cells at 32-cell and 64-cell stages. These gene states can be robustly predicted by the stemness GRN but not by randomized versions of the stemness GRN. Strikingly, we found a hierarchical structure of the GRN with the Oct4/Cdx2 motif functioning as the first decision-making module followed by Gata6/Nanog. We propose that stem cell populations, instead of being viewed as all having a specific cellular state, can be regarded as a heterogeneous mixture including cells in various states. Upon perturbations by external signals, stem cells lose the capacity to access certain cellular states, thereby becoming differentiated. The findings demonstrate that the functions of the stemness GRN is mainly determined by its well-evolved network topology rather than by detailed kinetic parameters.

# 1. Introduction

Embryonic stem cells (ESCs) can differentiate into cells of specialized types in a precise and organized manner, and dysregulation in stem cell differentiation results in early fetal death or severe disease [1–3]. Due to its essential role in survival for all multicellular organisms, stem cell differentiation must be highly conserved in order to allow for precise decisions at each step of lineage commitment. However, recent experimental results suggest that some transcription factors (TFs) such as Nanog exhibit heterogeneous expression levels at the single-cell level in mouse ESCs [4–8]. It remains largely unknown how the regulatory machinery of stemness performs its robust function in the presence of significant cell-to-cell heterogeneity. The answer to this question will shed light on the regulatory mechanism of stem cell differentiation, a crucial step toward better cellular reprogramming and stem cell-based therapies.

A substantial amount of research has been conducted to identify key TFs and their roles in directing stem cell differentiation [9–11]. These accumulated data enable us to map the underlying gene regulatory networks (GRNs). To elucidate the operating principles of these GRNs, computational approaches have been applied [12–16]. In particular, some of these computational studies adopted a bottom-up approach to construct gene regulatory networks (GRNs) with a small set of master regulators, and assume that the decision-making of stem cell differentiation is driven by the master regulators, such as the TFs Oct4, Sox2 and Cdx2. The dynamics of the GRNs can be simulated by either deterministic [17–19] or stochastic approaches [13,20–24]. These studies have indeed provided valuable insights into the regulatory mechanism underlying stem cell differentiation. However, these studies typically suffer from three issues. First, the modeling analysis typically focuses on only a standalone gene circuit, and therefore the effects of other genes and heterogeneous microenvironments cannot be included. Second, the exact values of kinetic parameters needed for modeling are largely unavailable, and since the modeling results depend on the estimated parameters, this issue can severely limit the predictive power of the models. Third, most studies do not provide a systematic way to quantify the robustness and plasticity of GRNs.

To address these issues, we here have applied a recently developed mathematical modeling algorithm, *r*andom *ci*rcuit *pe*rturbation (RACIPE) [25,26], to explore the robust dynamical behaviors of a proposed core GRN governing stemness (**figure 1**). RACIPE was developed to elucidate the robust gene expression patterns (also referred to as gene states) and generic features of transcriptional regulatory networks [25–27]. Unlike traditional approaches, RACIPE takes the topological information of a network as the only input, and generates an ensemble of mathematical models. Each mathematical model is simulated by the same set of chemical rate equations with different sets of kinetic parameters representing heterogeneous signaling and epigenetic states. The parameters of each model can differ by up to one or two orders of magnitude and are generated randomly under a specially designed sampling scheme (e.g. half-functional rule). Multiple initial conditions are used for each model in order to identify all possible steady-state solutions. The parameters and the corresponding stable steady-state solutions generated by the ensemble models are collected and subject to statistical analysis, by which the robust gene states are elucidated. It has been shown that RACIPE successfully identifies the gene states enabled by circuit motifs (i.e., toggle-switch-like circuit, repressilator, coupled toggle-switches) and GRNs governing epithelial-mesenchymal transition (EMT) and B cell development [26–28].

Here, we use RACIPE to analyze a core stemness GRN which contains eight master regulatory TFs involved in stem cell differentiation. We find that applying RACIPE to the stemness GRN can recapitulate the gene expression patterns of mouse ESCs at 32-cell and 64-cell stages. These gene expression patterns are conserved robust features of the stemness GRN but disappear when the network topology is randomized. Furthermore, through performing *in silico* perturbation analysis, we show that (a) the stemness GRN enables fifteen gene states and the presence of external signals can exclude the accessibility of some of these states; (b) the stemness GRN has a hidden hierarchical structure, which enables a two-step decision-making process with the Oct4/Cdx2 motif functioning as the first decision-making module and the Gata6/Nanog motif as the second one. In summary, we demonstrate that the robustness of stemness regulation is mainly determined by the topology of the stemness GRN, and RACIPE can be applied straightforwardly to elucidate the hierarchical decision-making of stem cell differentiation.

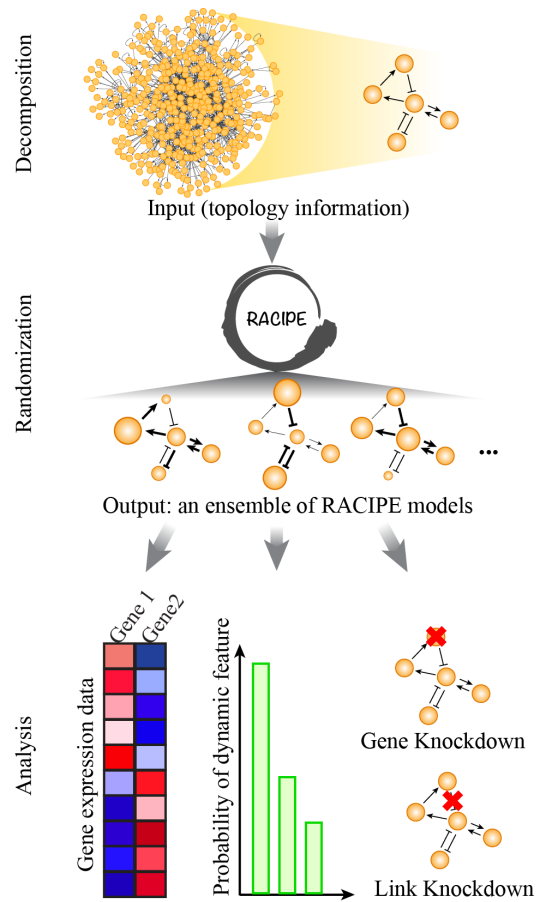

**Figure 1. Schematic illustration of _ra_ndom _ci_rcuit _pe_rturbation (RACIPE).** The gene regulatory network governing a specific cellular function can be divided into two parts - a core decision-making module and the rest functioning as input signals to the core. Through randomization, RACIPE generates an ensemble of mathematical models, each of which is simulated by the same set of chemical rate equations but with randomly sampled parameters. The simulation results of the model ensemble are subject to statistical analysis, such as hierarchical clustering analysis (HCA), and *in silico* gene/link perturbation analysis.

## 2. Materials and Methods

### 2.1. Mathematical modeling of the stemness GRN

In this study, the dynamical behavior of the stemness GRN (network diagram illustrated in figure 2a, details illustrated in section 3.1) was studied by RACIPE, Specifically, the RACIPE procedure creates an ensemble of models in each of which the temporal dynamics of the eight TFs (Oct4, Sox2, Cdx2, Gata6, Gcnf, Pbx1, Klf4 and Nanog) and the protein complex (Oct4-Sox2) are simulated by a set of ordinary differential equations (ODEs) accounting for their production, degradation and regulatory interactions. The transcriptional regulation between these TFs is modeled by the shifted Hill function [29]. The full details about the mathematical model and the implementation can be found in **electronic supplementary material, §S1 and Table S1**. To account for the binding/unbinding reactions between the TFs Oct4 and Sox2, which is not captured in the original RACIPE, we generalize the algorithm by modifying the rate equations to capture association and disassociation of the protein complex Oct4-Sox2. Full details regarding generalized RACIPE and its implementation can be found in **electronic supplementary material, §S2 and Table S2.**

### 2.2. Analysis of the RACIPE-generated gene expression data and experimental data

The details about the normalization of RACIPE-generated gene expression data can be found in **electronic supplementary material, §S3.** Clustering analysis has been performed on the normalized RACIPE-generated data to identify the gene expression patterns (**electronic supplementary material, §S4**). The details of the method to compare RACIPE-generated data with experimental data can be found in **electronic supplementary material, §S5-6.**

## 3. Results

### 3.1. RACIPE identifies robust gene states enabled by the stemness GRN

We integrate the master gene regulators of stemness characterized by previous studies [13,23,30], and construct a core stemness GRN. The GRN is composed of eight TFs (Oct4, Sox2, Cdx2, Gata6, Gcnf, Pbx1, Klf4 and Nanog) and one protein complex (Oct4-Sox2) (**figure 2a**). Due to the complexity of the GRN, elucidating its dynamical behaviors can be difficult through traditional modeling approaches. Here, we utilize RACIPE to identify the robust gene states enabled by the stemness GRN. As all regulatory links in the stemness GRN are transcriptional except for the binding/unbinding process between the TFs Oct4 and Sox2, for the simplicity, we initially model the temporal dynamics of the Oct4-Sox2 complex in the same manner as the other TFs. Later, we will generalize RACIPE to include binding/unbinding reactions and verify that none of the results change in any meaningful ways.

In our approach, instead of finding a representative set of kinetic parameters, we randomly generate 10,000 sets of parameters (i.e., 10,000 RACIPE models) within their given biologically reasonable ranges. For each model, we numerically solve the governing ODEs with 1,000 random initial conditions so as to thoroughly identify all possible stable steady-state solutions. These 1,000 initial conditions give rise to 1,000 solutions, based on which we identify the number of distinct stable states and their corresponding gene expression profiles. We show that 1000 initial conditions are sufficient, as increasing the number of initial conditions to 1,500 or 2,000 generates consistent probability distributions of the number of stable states (**figure 2b)** and stable states **(electronic supplementary material, figure S1**) of the 10,000 models.

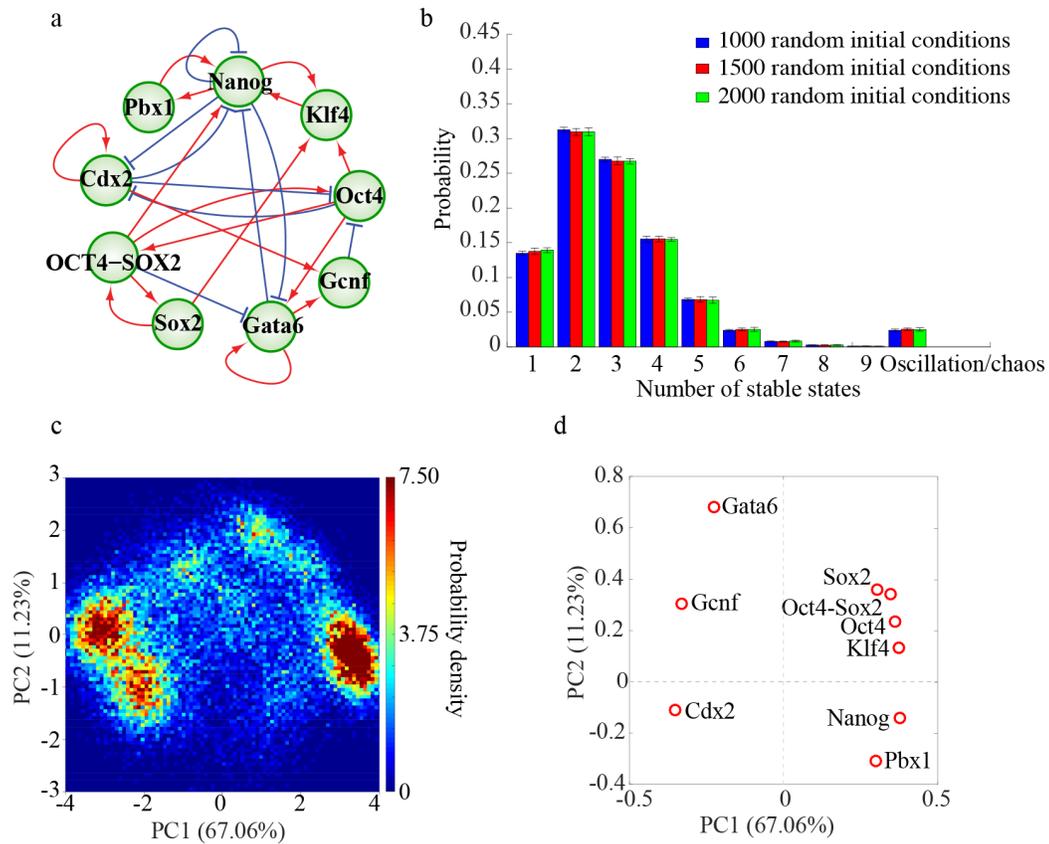

**Figure 2. The RACIPE method uncovers robust gene states allowed by the stemness GRN.** (a) Diagram of the core gene regulatory network governing stem cell differentiation. Red arrows represent excitatory regulation; blue bar-headed arrows represent inhibitory regulation. (b) Probability distribution of the number of stable steady states generated by 10,000 RACIPE models. Different colors represent different cases characterized by different numbers of initial conditions (blue: 1000 times, red: 1500 times, and green: 2000 times) that are used to simulate each RACIPE model. Each case was repeated 10 times to estimate the mean and the standard deviation of the distribution. (c) 2D probability density map of the RACIPE-predicted gene

expression profiles projected onto the 1$^{st}$ and 2$^{nd}$ principal component (PC1 and PC2) axes. (d) Contribution of each gene to PC1 and PC2. The PCs were obtained by performing principal component analysis (PCA) using the gene expression profiles from all 10,000 RACIPE models.

For the majority of RACIPE models (~98%), the stemness GRN allows one to six stable steady states (**figure 2b**). There are rare occasions (<1%) where RACIPE models generate more than six stable steady states. Since they are not statistically significant, we excluded these data for further analysis. The circuit also has ~ 2% chance of having oscillatory or chaotic dynamics (e.g., time-dependent dynamics), which are not the focus of this manuscript and therefore are being excluded for further analysis. We collected the gene expression profiles from all 10,000 RACIPE models and constructed a data matrix, where each column represents a gene and each row represents a stable steady-state solution. These data resemble experimental gene expression data, thus inspiring us to apply similar statistical methods.

Since the kinetic parameters of the circuit are randomized with large variations (up to one or two orders of magnitude), one might expect that the gene expression profiles from different models would be very different. Strikingly, we found that the gene expression profiles can be segregated into only a few clusters when projected onto two independent components by the commonly used principal component analysis (PCA) (**figure 2c**). Regarding the first principal component (PC1), Oct4, Sox2 and Nanog contribute positively while Cdx2 and Gata6 contribute negatively (**figure 2d**), indicating an anti-correlation of the activity between these two sets of TFs, which is consistent with the experimental observations [31]. We have also shown that different ranges for parameter randomization and different types of random sampling distribution (Uniform or Gaussian) in RACIPE all generate largely consistent probability density maps (**electronic supplementary material, figures S2-3**), as compared with the one shown in **figure 2c**.

## 3.2. RACIPE-generated gene expression profiles are consistent with experimental observations and match single-cell gene expression data of mouse ESCs at 32-cell and 64-cell stages

To identify the pattern of RACIPE-generated gene expression profiles, we applied HCA to the RACIPE-generated gene expression data. We found that the RACIPE-generated data form fifteen major clusters, representing fifteen different gene expression patterns (**figures 3a, b**). We found that most of the TFs exhibit bi-modal distributions, which indicates the up-/down-regulation of these TFs is associated with different phenotypes during stem cell differentiation (**figures 3a, electronic supplementary material, figure S4**). To evaluate how well RACIPE can recapitulate the characteristic gene expression of various cellular phenotypes during stem cell differentiation, we compared the RACIPE-generated gene expression profiles with those observed experimentally. We found that the RACIPE-generated gene states recapitulate the gene expression patterns measured by experiments (**figure 3c, electronic supplementary material, figure S5**). We compare the RACIPE-generated data with the gene expression data of single mouse embryo cells at various stages during development [32] (**figure 3a, electronic supplementary material, §S6, figures S6-8**). Interestingly, the RACIPE-generated gene states match those during the late stage of embryo development (≥32 cells), where totipotent cells start to differentiate into trophectoderm (TE) and inner cell mass (ICM) [33], but not those from the early stage (≤16 cells). The experimental verified gene states - $Cdx2^{Hi}$, $Gata6^{Hi}/Nanog^{Hi}/Oct4^{Hi}/Sox2^{Hi}$, $Nanog^{Hi}/Oct4^{Hi}/Sox2^{Hi}$ and $Gata6^{Hi}/Oct4^{Hi}/Sox2^{Hi}$ - are among the most probable gene states identified by RACIPE (**figures 3b**). Notably, the matching between RACIPE data and experimental data is statistically significant (**electronic supplementary material, figures S7-8**).

Indeed, the RACIPE-generated gene expression patterns can characterize multiple developmental stages. For example, the first cell fate determination during embryonic development happens at the blastocyst stage when the ICM and TE are formed [33]. Oct4 and Sox2 were reported to be expressed throughout ICM (**State 1,** the state numbers are indicated in **figure 3c**) [34,35]. At the early stage of ICM differentiation, Gata6 and Nanog exhibit co-expression (**State 6**) [36,37], but Nanog, Oct4 and Sox2 are required for cells to commit to epiblast and reach the ground state of pluripotency (**States 1 and 12**) [34,35]. Further differentiation of mESCs into mesendoderm requires Nanog and Oct4 but not Sox2 (**State 8**) [38]. Upon Gata6 induction, mESCs exhibit a step-wise pluripotency factor disengagement, starting with down-regulation of Nanog and Esrrb, then Sox2, and finally Oct4, along with a step-wise coexpression of extraembryonic endoderm (ExEn) genes (**State 5**) [35]. On the other hand, down-regulation of Oct4 induces the differentiation of mESCs into trophoblast characterized with high expression of Cdx2 and Gata6 (**State 2**) [30,39]. Overexpression of Cdx2 is sufficient to generate proper trophoblast stem cells (**State 3**) [31,32]. TE could further differentiate into ExEn, where Cdx2, Gata6 and Sox2 can be all expressed (**States 7, 9, 11 and 15**) [31,40,41]. A summary of the correspondence between RACIPE-generated gene states and experimental observed gene expression profiles during development can be found in **electronic supplementary material, figures S5.**

These results suggest that RACIPE can identify the gene expression patterns of various cellular phenotypes especially those of the late developmental stages and also characterize potential additional gene states during stem cell differentiation.

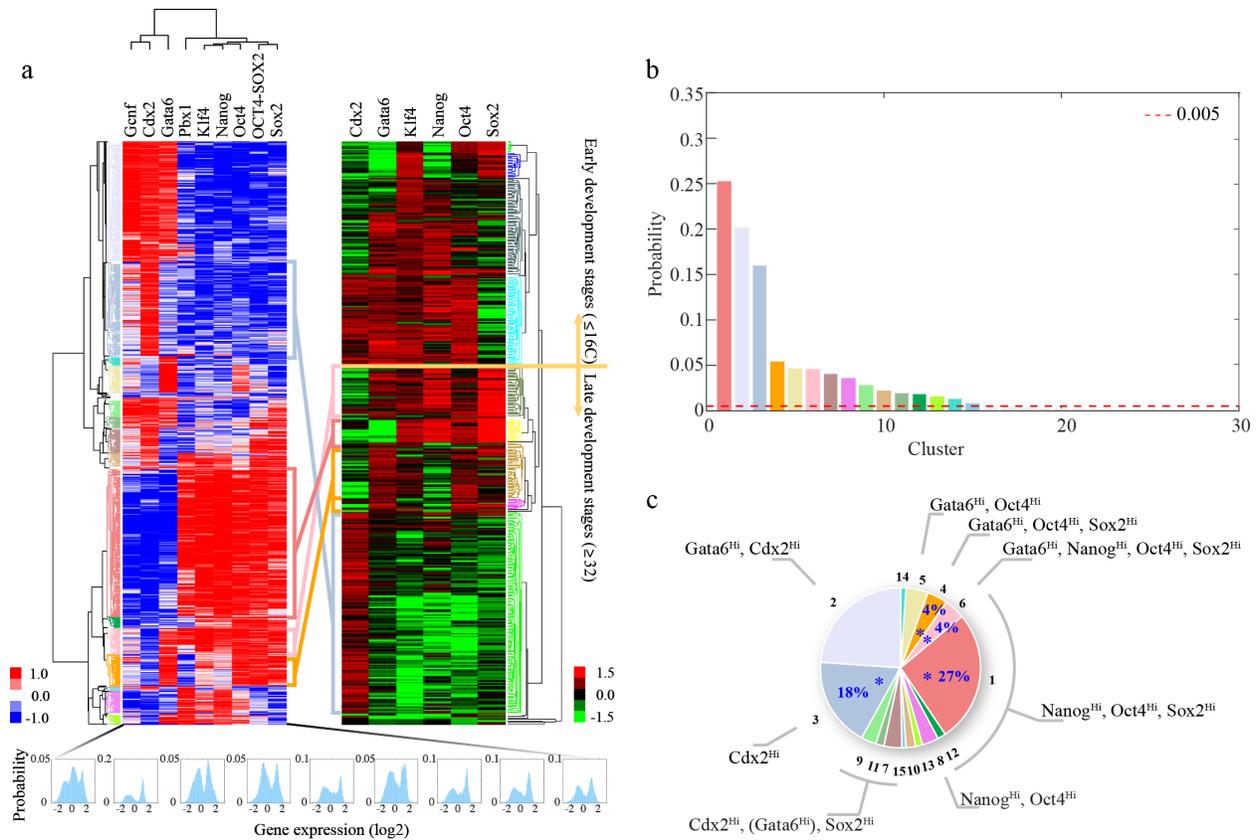

**Figure 3. Comparison of the RACIPE-generated gene expression profiles and single-cell gene expression data of mouse embryo**. (a) Robust clusters (gene states, colored hierarchical trees) were identified for both data sets by unsupervised HCA. Four RACIPE-predicted gene states match those from the late stage single-cell gene expression data. The histogram of the predicted expression levels for each gene is shown at the bottom (blue, 50 bins in each histogram). In both heat maps, each column represents a gene; each row represents the gene expression profile of a stable steady state of a RACIPE model (left) or that for a single cell

(right). (b) A total of 54 gene clusters (only show 30 here) were identified by HCA. With a minimum probability cutoff of 0.005, we identified 15 clusters, referred to as major gene states. The coloring scheme for these 15 clusters is consistent with that used in (a), and the other clusters are shown in grey. (c) The characteristic gene expression of each gene state ranked by the likelihood in the RACIPE models. The four gene states that match the experimental data are highlighted by blue asterisks and are shown with their likelihoods (The method to classify the gene states in the presence of external signal can be found in **electronic supplementary material, §S8**).

### 3.3 The topology of the stemness GRN determines the robust gene states

To further investigate the role of the topology of the stemness GRN in maintaining the robust gene states, we compare the stemness GRN with two types of its randomized versions (Type I including 10 networks and Type II including 10 networks) with randomly generated connections among genes. Both Type I and Type II randomized networks preserve the total number of inward and outward links for each gene, and the binding of Oct4 and Sox2 (**electronic supplementary material, fig. S9**). For the Type I randomized network, we also keep the same numbers of excitatory and inhibitory inward links for each gene. We then apply RACIPE to these two types of randomized networks and compare their dynamic behaviors with those of the actual stemness GRN.

Neither Type I nor Type II randomized networks can generate the aforementioned robust gene states and recapitulate the experimentally observed gene expression features. Compared to the stemness GRN, both Type I and Type II randomized networks are much more likely to generate oscillatory or chaotic dynamics but not multi-stable states for each RACIPE model. Instead, both randomized networks are more likely to have only one stable state for each RACIPE model (**electronic supplementary material, figure S10**). When the gene states from all RACIPE models are combined, the histogram of each gene expression generated by the stemness GRN typically exhibits multi-modal distributions but those generated by the randomized networks mainly exhibits mono-modal distribution (**electronic supplementary material, figure S11**). In addition, we also find that it is difficult to cluster the gene expression data generated by the randomized networks (**electronic supplementary material, figure S12**), partly because the stemness GRN has much higher local density of the RACIPE-generated gene expression data relative to the randomized networks (**electronic supplementary material, §S7, figure S11**)..

Furthermore, the RACIPE-predicted gene expression profiles of the stemness GRN are significantly better than those of the randomized networks in recapitulating the experimental observation of the gene expression features of various stages during development, and especially the single-cell gene expression data of mouse embryo (**figure 4 and electronic supplementary material, figure S13**). In summary, these results indicate that the topology of the stemness GRN has been well-evolved to be robust in regulating stem cell differentiation in the presence of perturbations.

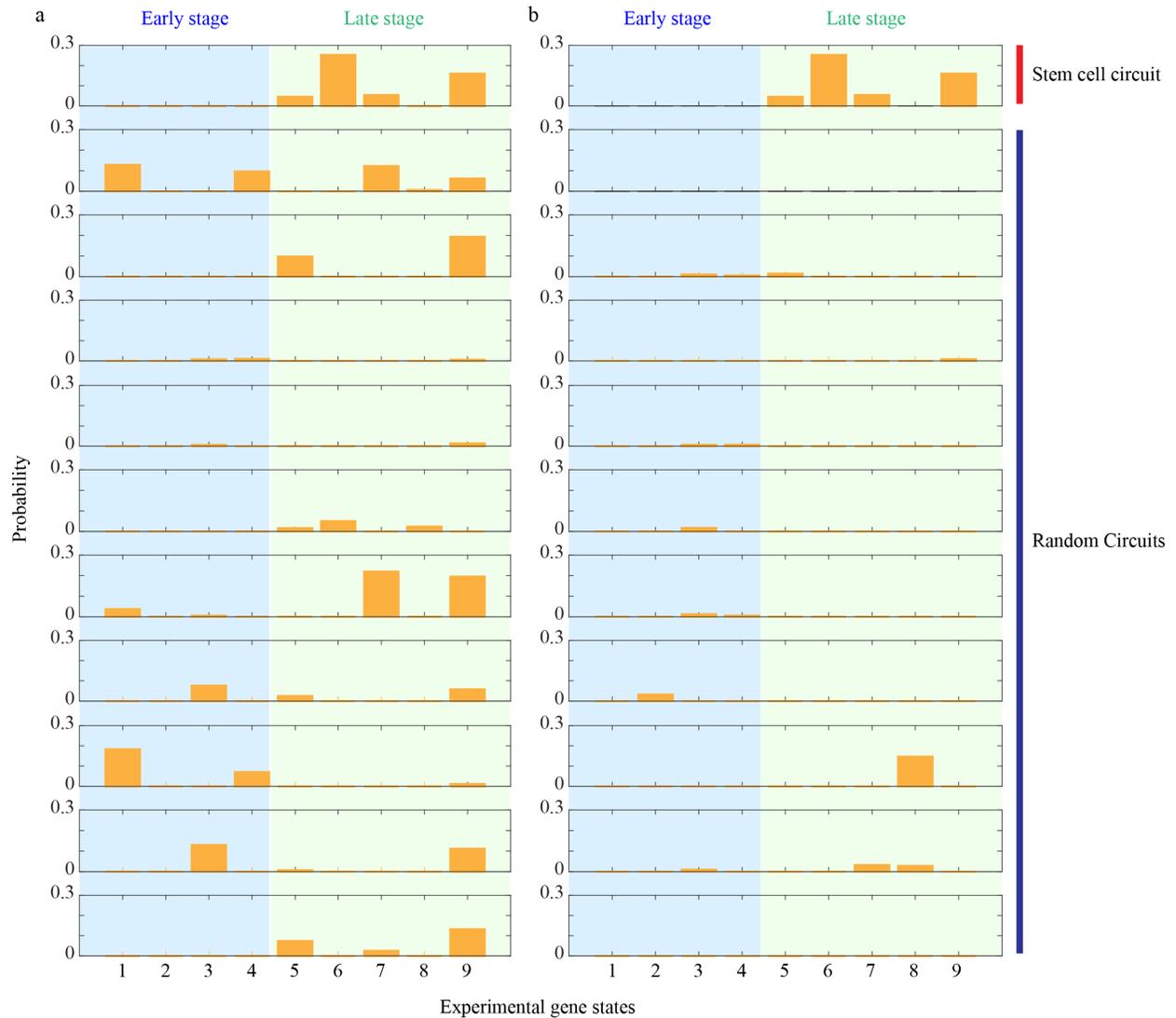

**Figure 4. Comparison between the stemness GRN and the randomized networks (10 for Type I (a) and 10 for Type II(b)).** Percentage of the RACIPE predicted gene expression data matching each experimental gene state shown in Fig. 3a (right) for the stemness GRN and

random networks. The details of the 10 Type I randomized networks and the 10 Type II networks can be found in **electronic supplementary material, figure S11**.

**3.4 RACIPE identified the key parameters representing key stimuli that promote transitions between different phenotypes during stem cell differentiation**

In addition to characterizing the gene expression profiles of different cell phenotypes, RACIPE enables us to uncover the key parameters representing physiological conditions that mediate the phenotypic transitions during stem cell differentiation. From the output of RACIPE simulations, we can identify the kinetic parameters that are significantly changed between gene states (**electronic supplementary material, §S7, figure 5a**). We will discuss two examples of this feature in the following paragraphs.

To characterize the most differential kinetic processes between gene state 1 (representing the pluripotent epiblast stage) and gene state 2 (representing the trophoblast stage), we quantify the change of the mean values of each parameter in gene state 2 relative to gene state 1. We identify the parameters whose values increase the most (e.g., the degradation rate of Sox2 and the degradation rate of Oct4-Sox2), and the parameters whose values decrease the most (e.g., the production rate of Sox2, the degradation rate of Gata6 and the threshold of Gata6 self-activation) in gene state 2 relative to gene state 1 (**figure 5b**). The results indicate the transition from gene state 1 to gene state 2 is characterized by increased degradation of Sox2, decreased production of Sox2 and accumulation of Gata6. This indication is consistent with the experimental result that loss of Sox2 results in differentiation of ESCs as characterized by the up-regulation of trophoblast markers [42].

To characterize the stimuli that can promote the transition from gene state 1 (representing the pluripotent epiblast stage) to gene state 6 (representing the early stage of ICM differentiation), we quantify the changes of each parameter in gene state 6 relative to gene state 1. We identify the parameters whose values increase most (the maximum production rate of Gata6) and the parameters whose values decrease most (the threshold for Gata6 self-activation and fold-change of the inhibition of Gata6 by Nanog) in gene state 6 relative to gene state 1 (**figure 5a, c**). The results indicate that transition from gene state 1 to gene state 6 requires up-regulation of Gata6 that can be accomplished by either increasing Gata6 production or decreasing the inhibition of Gata6 by Nanog. This simulation result is again consistent with the experimental observation that Gata6 is highly expressed at the early stage of ICM differentiation [36]. The full details regarding the key parameters shifts between any two of the gene states, as shown in **figure 5a**, can be found in **electronic supplementary material, figure S14.**

In summary, RACIPE can identify the parameters that differ the most between different gene states, such as the production/degradation rates and the threshold/fold-change of a regulatory link. Consequently, the biological process represented by these parameters can be considered as the regulatory target needed to drive cells to undergo certain phenotypic transitions.

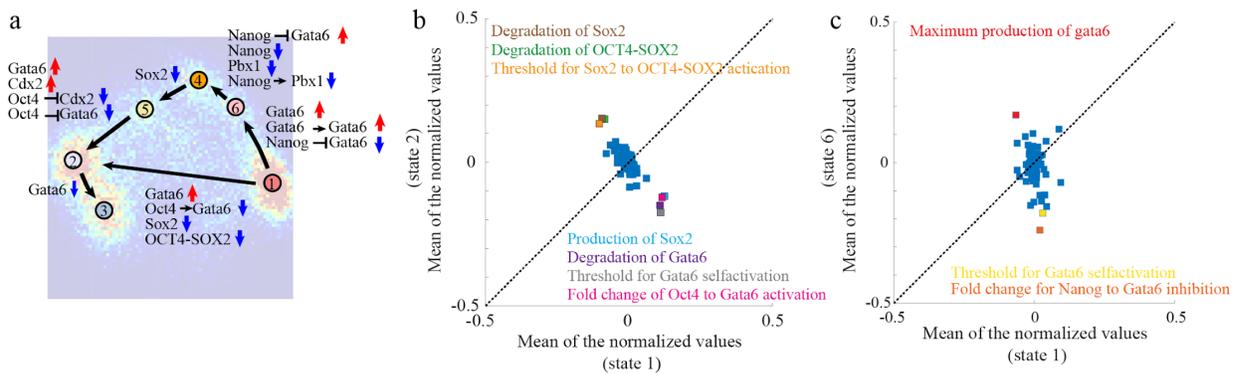

**Figure 5. Key parameters that are involved in the transitions among certain gene states.** (a) A summary of the results depicted on top of the probability density map **(figure 2a)** of the RACIPE-generated gene expression data. In (a), along with each transition the key parameters that have shifted the most have been marked. Red up arrows represent up-regulation and blue down arrows represent down-regulation. The mean of the normalized values for each parameter for the two corresponding gene states (x-axis for the first gene state, and y-axis for the second gene state) are shown in (b) (states 1 and 2) and (c) (states 1 and 6). The change of the parameters values between any of the transitions shown in (a) can be found in **electronic supplementary material, figure S7**.

### 3.5 The stemness GRN exhibits a hierarchical decision-making structure

RACIPE also enables in silico perturbation analysis of the stemness GRN, including knocking out genes and removing links, by which we can understand the role of the knocked-out gene or the removed link in the dynamical behaviors of the stemness GRN. Here, we perform two types of perturbation analyses, knocking out a gene each time or removing a regulatory link each time (**electronic supplementary material, §S8, 9**). In both types of perturbation analyses, we quantify the change of the probability distribution of the multi-stability exhibited by the stemness GRN using the Kullback–Leibler (KL) divergence (**electronic supplementary material, §S10**).

Through analyzing the gene knockout results, we found that the knockout of TFs Cdx2, Oct4, Sox2 and the complex OCT4-SOX2 leads to the most significant changes in the probability distribution of the multi-stable behavior of the GRN (**figure 6a**). Strikingly, removal of the regulatory links among these specific TFs also leads to the most significant changes in the probability distribution of the number of stable states (**figure 6b**). These TFs and the regulatory links between them indeed form a sub-network, representing the first decision-making module, referred to as the Oct4/Cdx2 module (**figures 6c, d**). The rest of the TFs and regulatory links form a sequential second sub-network, referred to as the Gata6/Nanog module (**figure 6d**). The RACIPE simulation results are consistent with experimental observations showing that the TFs Oct4 and Cdx2 govern the commitment of totipotent cells to either the ICM (Oct4$^{high}$) or the TE (Cdx2$^{high}$) lineages and the Gata6/Nanog module governing the commitment of ICM cells to either epiblast (Nanog$^{high}$) or primitive endoderm (Gata6$^{high}$) lineages [43].

To compare the behaviors of these two sub-networks, we apply RACIPE to each of them and then compared the RACIPE-generated gene expression profiles with those generated by applying RACIPE to the full network. In other words, we want to evaluate how the dynamic behaviors of the two sub-networks change upon removal of the regulatory links connecting them. We find that the gene expression profiles determined by the first decision-making module Oct4/Cdx2 is conserved while those determined by the second module Gata6/Nanog are largely disrupted, upon the removal of the regulatory links connecting these two modules (**electronic supplementary material, figure S15**). The results here support the hierarchical structure of these two decision-making modules.

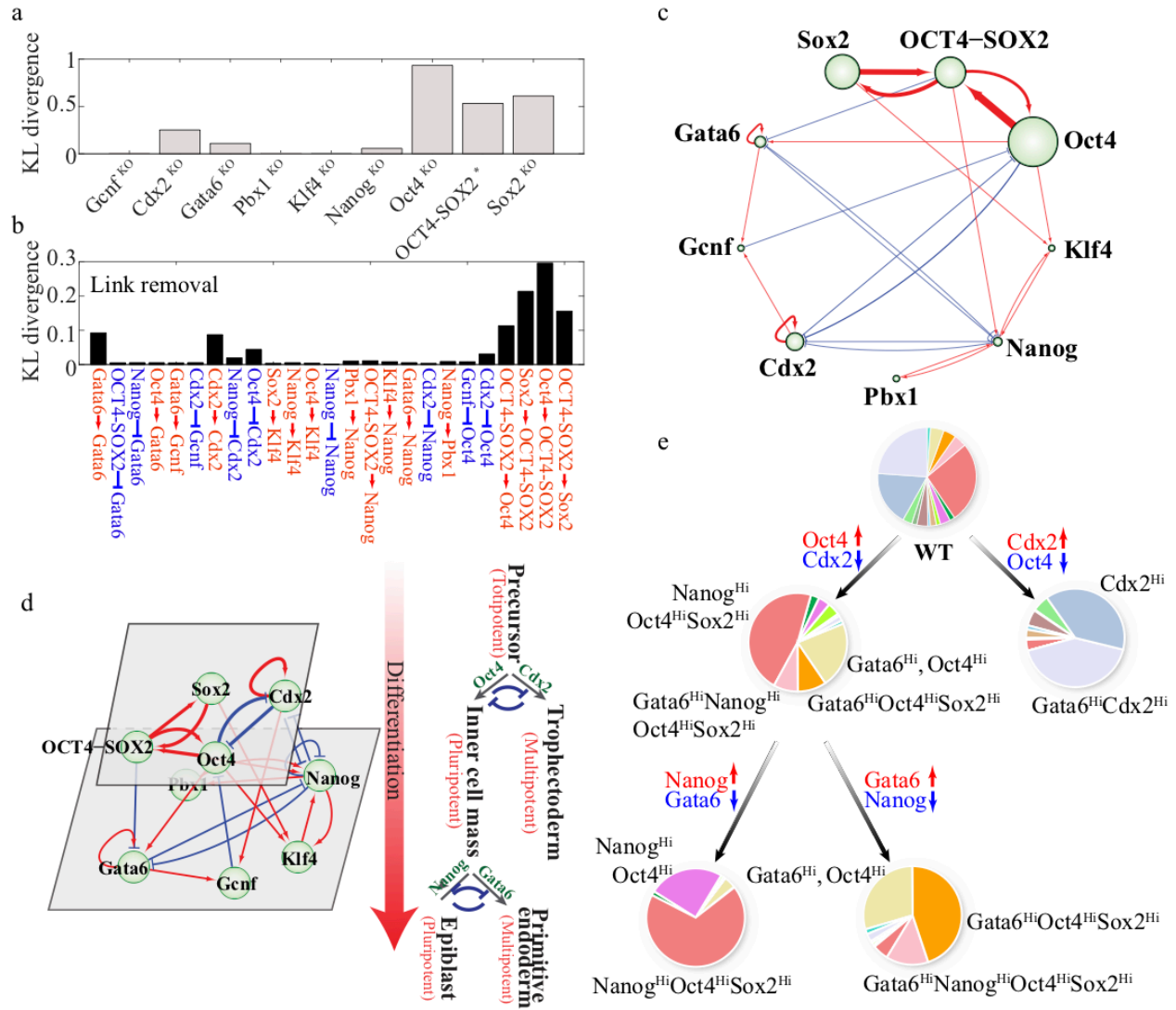

**Figure 6. Hierarchical structure of the stemness GRN inferred from the perturbation analysis.** (a) The Kullback-Leibler (KL) divergence between the probability distributions of the number of stable states for each RACIPE model computed before and after the knockout (KO) of each gene. "Oct4-Sox2*" represents the removal of the protein complex Oct4-Sox2. (b) Similar to (a), but the KL divergences are between the distributions before and after removal of each regulatory link. (c) Schematic diagram of the stemness GRN highlighting the important genes and regulatory links. The larger the gene element and the thicker the regulatory link, the more important the component to the network behavior, as inferred from the analyses in (a) and (b). (d) The hierarchical structure of the stemness GRN (left) is consistent with the two-step decision-making of mouse embryonic development (right). (e) The roadmap of stem cell differentiation inferred from the RACIPE simulations. All the original RACIPE models (WT) were treated by activating (↑) or inhibiting (↓) the maximum production rate of the corresponding genes by 50-fold. The probability of different gene states is proportional to the area in the pie chart.

## 3.6 Gene states of the stemness GRN can become inaccessible upon external signals

As stem cell differentiation is a cascading event induced by signaling, we would like to analyze how the stemness GRN responds to various external signals (e.g., signals acting on the TFs in the network but getting no feedback from the circuit) by RACIPE. The effects of external signals on a certain gene are simulated by scaling the production rate of that gene to 50-time larger (representing excitatory signal) or smaller (representing inhibitory signals). We then calculated the probability distribution of gene states upon the imposition of external signals (excitatory or inhibitory) on each gene. As we showed before, without any external signals, the stemness GRN allows 15 robust gene states, referred to as the wild type (WT) (**figure 6e**). Relative to the WT, up-regulation of Cdx2 restrict most RACIPE models to acquire the stable states with $Cdx2^{Hi}$ (representing the trophectoderm stage) [31], while up-regulation of Oct4 restrict most RACIPE models to acquire the states with $Oct4^{hi}$ (representing the ICM stage) [36,43] (**figure 6e**). After cells reach the ICM stage, additional signals acting on Nanog and Gata6 can convert the GRN largely to either the $Nanog^{Hi}$ state (representing the epiblast stage) or the $Gata6^{Hi}$ state (representing the primitive endoderm stage) [43] (**figure 6e**). These simulation results indicate that external signals often do not create new gene states instead make a subset of gene states more accessible and the rest less accessible as also observed in **figure 5** in our previous work [26]. In other words, a stepwise administration of external signals is able to restrict the gene expression of the stemness GRN to specific cellular states.

With the additional assumption that different sets of kinetic parameters represent different single cells, we can regard the ensemble of RACIPE models as an approximation to the heterogeneous cell population, where each cell is defined by a distinct set of parameters. We can then quantify the effect of various external signals on population heterogeneity using information entropy theory (**electronic supplementary material, §S16**). We systematically simulated the external signals acting on each TF by scaling the maximum production rate of that TF from 1/100 to 100 of the base level, representing inhibitory signals and excitatory signals respectively. For each TF, we apply RACIPE to the stemness GRN considering 20 different scenarios representing 10 inhibitory and 10 excitatory signals with varying strengths on that TF. We then applied the entropy-based index to quantify the population heterogeneity of each scenario for each TF

**(electronic supplementary material, figure S9)**. We show that the WT stemness GRN exhibits the highest entropy, e.g., highest heterogeneity, and external signals that either up-regulate or down-regulate the maximum production rate of a TF usually decreases the entropy, leading to decreased heterogeneity of the cell population. The decrease of population heterogeneity partially results from the limited access to only a few gene states instead of all due to external signals. Altogether, our results suggest that RACIPE can explore the possible roles of external signals in the dynamic behaviors of the stemness GRN, and these external signals may indeed restrict the gene states that can be accessible.

**3.7 Toward generalizing RACIPE by including the binding/unbinding details**
RACIPE provides a straightforward way to identify the robust dynamic behaviors of the stemness GRN. As we discussed before, RACIPE was originally developed for transcriptional regulation. In the stemness GRN, in addition to the majority of the links representing transcriptional regulation, there is one binding/unbinding process, between the TFs Oct4 and Sox2. We have therefore extended the RACIPE framework to explicitly model this binding/unbinding process to analyze how that may affect the network behavior. We performed a parallel analysis of the stemness GRN using the updated RACIPE including binding/unbinding details (referred to as RACIPE-wb) **(figure 7a, electronic supplementary material, figures S17, 18)**. We observed consistent gene states by RACIPE-wb **(figure 7b, electronic supplementary material, figures S19-21)** relative to those acquired by RACIPE **(figure 2c)**. We show that the RACIPE-wb generated gene expression profiles are quantitatively consistent with the RACIPE-generated ones **(figure 7c)**. By RACIPE-wb, we performed perturbation analysis by knocking out genes and removing regulatory links one by one. Consistent with the result by RACIPE **(figures 6a, b)**, we found that knocking out the TFs Oct4, Sox2, Cdx2 or Oct4-Sox2 or removing the regulatory links among these genes have the most pronounced effects on the multi-stable behaviors of the stemness GRN **(figures 7d-e)**. Indeed, RACIPE-wb amplifies the differences observed under link removal relative to RACIPE **(figures 6 and 7)**. Specifically, blockade of the binding between Oct4 and Sox2 (KL divergence = 0.32) in RACIPE-wb is so pronounced that it is equivalent to a full removal of the protein complex Oct4-Sox2 in the stemness GRN (KL divergence = 0.32). Blockade of the unbinding of Oct4-Sox2 (KL divergence = 1.57) in RACIPE-wb is so pronounced that it is approximately equivalent to knocking out both Oct4 (KL divergence = 0.95) and Sox2 (KL divergence = 0.64) (**figure 7d**). The effects of link removal for the rest of the regulatory interactions are in general smaller in RACIPE-wb than those in the original RACIPE. But the overall trends are similar for both modeling algorithms. Among these interactions, we found that removal of the inhibitory link from Cdx2 to Oct4 exhibits the highest KL divergence **(electronic supplementary material, figure S22)**. The result by RACIPE-wb also indicates a similar hierarchical decision-making structure of the stemness GRN with the Oct4/Cdx2 module forming the first decision-making module and the rest forming the second decision-making module. In summary, when the detailed binding/unbinding processes are considered, the RACIPE-wb characterized dynamic behaviors of the stemness GRN remain consistent with RACIPE characterized ones.

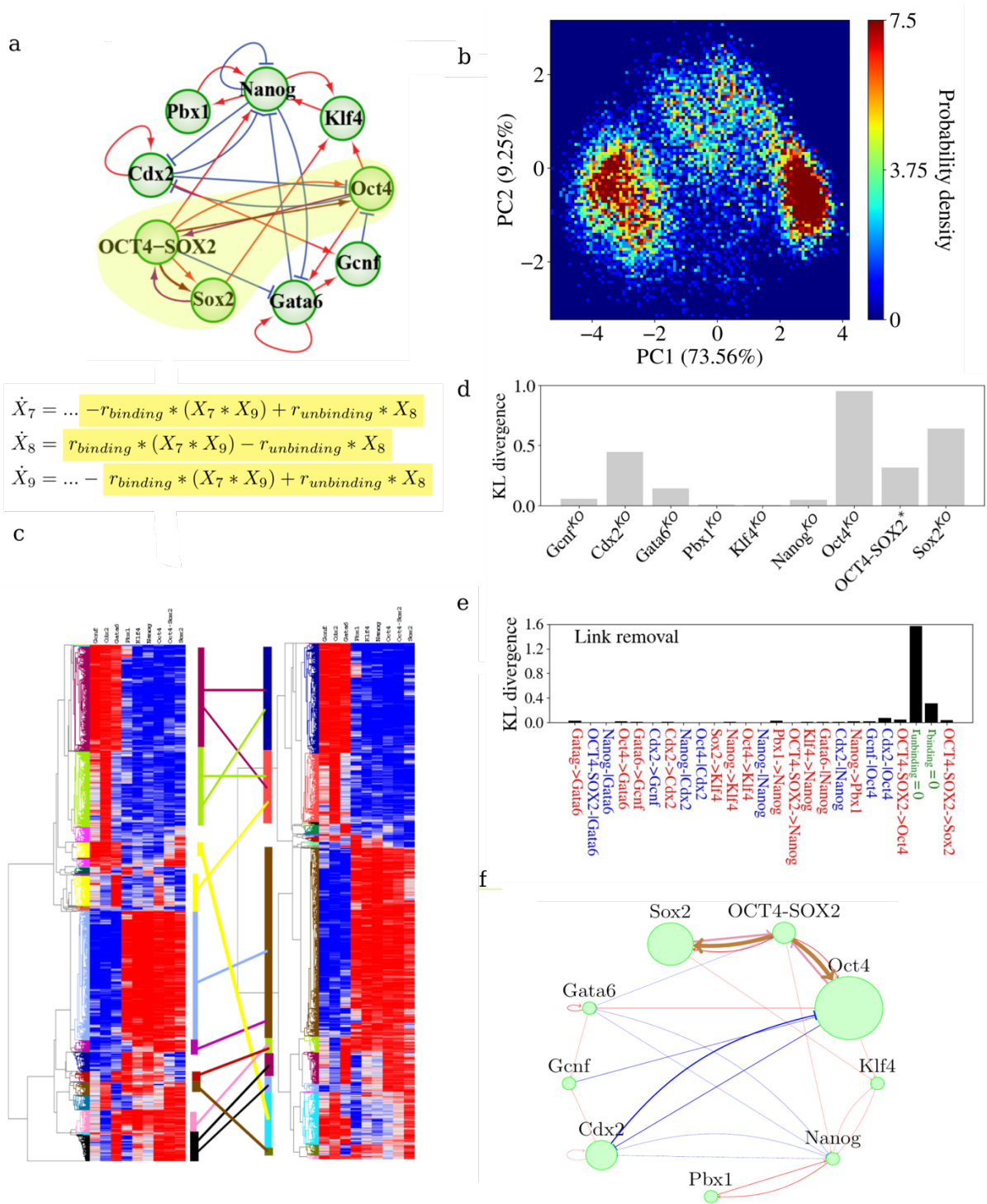

**Figure 7. Analysis results elucidated by applying the extended RACIPE (RACIPE-wb) to the stemness GRN reveals consistent characterization.** (a) Top panel: Diagram of the core stemness GRN highlighting the binding interactions between Oct4, Sox2, and the OCT4-SOX2 complex. Bottom panel: The main changes in the mathematical equations simulating the dynamics of Sox2 ($X_9$), Oct4 ($X_7$), OCT4-SOX2 ($X_8$) to capture their binding/unbinding

interactions. The full equations for RACIPE-wb are listed in **electronic supplementary material, §2**. (b) The 2D probability density map of the results for the RACIPE-wb model projected onto the first two principal components. (c) Unsupervised HCA for RACIPE and RACIPE-wb, left and right respectively. Clusters were identified using a probability cutoff of 0.005. The lines between clusters show the majority of clusters from the original RACIPE are also present in RACIPE-wb. Additionally, some of the clusters obtained by using RACIPE-wb were seen to be over/underrepresented as compared to results using the original RACIPE framework; the number of solutions belonging to that cluster is shown by the colored vertical bars on the left and right of the middle which correspond the dendrograms of HCA for RACIPE and RACIPE-wb, respectively. (d) The KL divergence of RACIPE-wb distributions before and after knocking out a gene. "Oct4-Sox2*" represents the removal of the protein complex Oct4-Sox2. (e) The KL divergence of RACIPE-wb distributions before and after the removal of a regulatory link. Also included is the blocking of binding between Oct4 and Sox2 and the blocking of the unbinding of OCT4-SOX2. (f) A schematic diagram depicting the relative importance of each gene and link as inferred by the analysis in (d-e).

# 4 Discussion

By applying RACIPE to the stemness GRN, we have identified fifteen distinct gene states. We found that the external signals acting on the stemness TFs can restrict the stemness GRN to acquire only certain gene states corresponding to differentiated cell phenotypes. The result may indicate a different interpretation of the Waddington's epigenetic landscape [44,45] for stem cell differentiation, in which people usually consider each progenitor and differentiated phenotype as a unique cellular state with distinct gene expression pattern, and cell differentiation as the transition from the progenitor state to a differentiated state. Our RACIPE simulation results indicate an alternative interpretation, where a stem cell population, instead of consisting of cells in a specific "stemness" state, can instead be regarded as a heterogeneous population of cells in various states, each of which corresponds to a differentiated lineage with a distinct gene expression pattern (**figure 8**) [46]. Cells with high cell potency are plastic and able to convert into the various cell states stochastically by both the intrinsic factors (gene expression noise, a fast process) and the extrinsic factors (transient epigenetic regulation and cell signaling, a slow process). However, when cells are subject to stable perturbations by external signals, they lose the capacity to access certain cellular states, therefore making the population less heterogeneous, i.e. smaller information entropy, and differentiated. Our view is consistent with the observation in experiments that the stem cell progenitors of either totipotency or pluripotency have highly heterogeneous gene expression, and several cell sub-populations of differentiated types, called lineage priming, have been identified in cell culture [47–49].

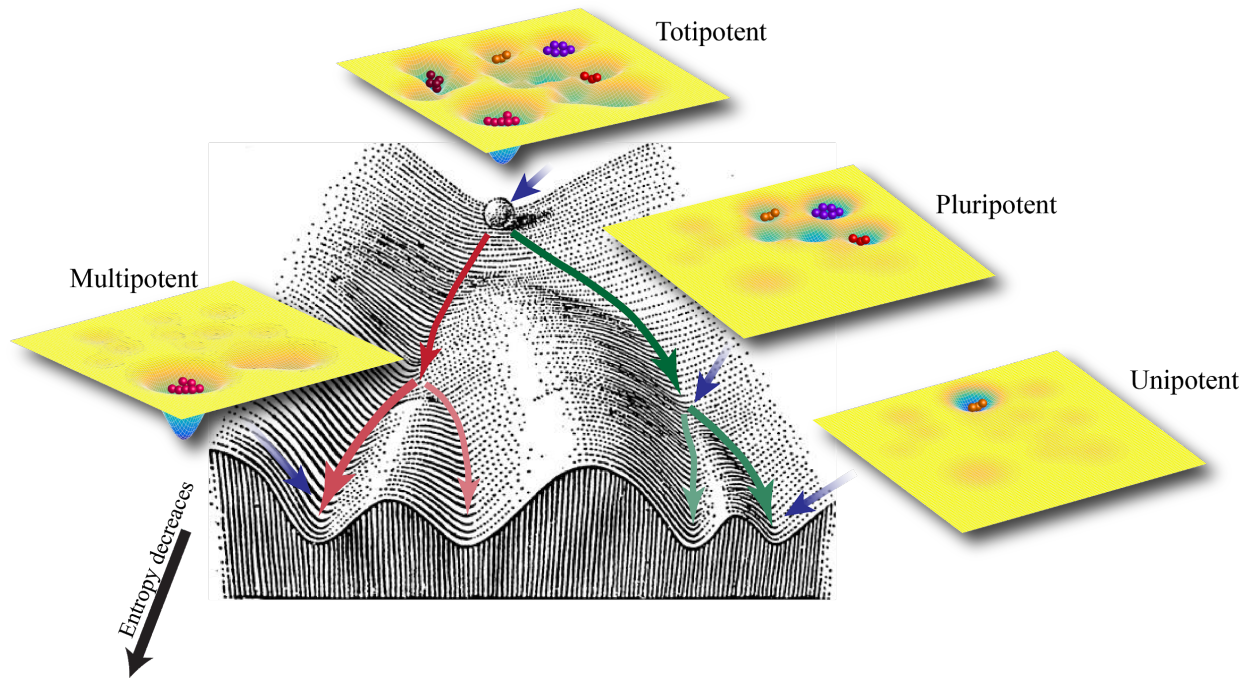

**Figure 8. Schematic illustration of the revised Waddington's epigenetic landscape for stem cell differentiation.** For each cell potency, the accessible cell types are shown by the attractors. Stem cells are induced by external signals toward differentiation along the valleys (highlighted by arrow lines with different colors) with the decrease of cell potency.

We showed that RACIPE-generated gene expression patterns recapitulate mainly the gene expression profiles of mouse embryos at late developmental stages (⩾32 cells) while not early stages. This can be due to the incompleteness of the decision-making network of stemness. Although the TFs in the stemness GRN have been shown to be the master regulators governing stem cell differentiation by experimental studies, it is possible that there are other important molecular regulators which are not included here. In other words, the quality of the stemness GRN can be the limitation of the current study. An experimentally validated GRN could be constructed by combining genomics data such as ChIP-Seq with biochemistry experiments [50–52]. However, it still remains a big challenge to construct reasonable large GRNs.

To conclude, by applying RACIPE to a core stemness GRN, we showed that the network topology plays an essential role in cell fate decision-making during stem cell differentiation. This result is analogous to the findings from protein structure modeling, where conformational motions have been found to be determined by the overall molecular shape [52] and protein folding process by native residue contacts [52,53]. RACIPE allows the interrogation of the robust dynamical behaviors of GRN by parametric randomization, from which we can identify the operating principles underlying the GRN functions.

**Acknowledgements:** This work was supported by National Science Foundation (NSF) Center for Theoretical Biological Physics (NSF PHY-1427654) and NSF grants DMS-1361411 and CHE-1614101. D.J. is supported by a training fellowship from the Gulf Coast Consortia, on the Computational Cancer Biology Training Program (CPRIT Grant No. RP170593). M.L. is supported by a startup fund from The Jackson Laboratory, by the National Cancer Institute of the National Institutes of Health under Award Number P30CA034196, and by the National Institute Of General Medical Sciences of the National Institutes of Health under Award Number R35GM128717. M.G. is supported by the NSF GRFP Number 1842494. This work was supported in part by the Big-Data Private- Cloud Research Cyberinfrastructure MRI-award funded by NSF under grant CNS-1338099 and by Rice University.